\documentclass[amsmath,comment,amssymb,twocolumn,showpacs,superscriptaddress,citeautoscript,prb]{revtex4}
\usepackage{graphicx}
\usepackage{amssymb}
\usepackage{epsfig}
\usepackage{dcolumn}
\usepackage{bm}
\usepackage{color}

\usepackage{amsmath, amssymb, latexsym, amscd, amsthm,amstext}
\usepackage{graphicx}
\usepackage{amsmath,amssymb}

 \usepackage{epstopdf} 

\newcommand{\tb}[1]{\textbf{#1}}
\newcommand{\be}{\begin{equation}}
\newcommand{\ee}{\end{equation}}
\newcommand{\bea}{\begin{eqnarray}}
\newcommand{\eea}{\end{eqnarray}}

\newcommand{\nn}{\nonumber}

\begin{document}
\title{Magneto-oscillations of the mobility edge in Coulomb frustrated bosons and fermions}

\author{Thuong T. Nguyen}
\affiliation{International School for Advanced Studies (SISSA), 34136 Trieste, Italy} 
\affiliation{The Abdus Salam International Center for Theoretical Physics, 34151 Trieste, Italy}
\affiliation{Istituto Nazionale di Fisica Nucleare (INFN), Sezione di Trieste, Trieste, Italy }
\author{Markus M\"uller}
\affiliation{The Abdus Salam International Center for Theoretical Physics, 34151 Trieste, Italy}
\affiliation{Condensed matter theory group, Paul Scherrer Institute, CH-5232 Villigen PSI, Switzerland}
\affiliation{Department of Physics, University of Basel, Klingelbergstrasse 82, CH-4056 Basel, Switzerland}

\date{\today }

\begin{abstract}
We study the crossover from strong to weak localization of hard-core bosons on a two dimensional honeycomb
lattice in a magnetic field, as motivated by recent experiments on structured films.
Taking into account long range Coulomb interactions among the bosons, an effective mobility edge in the excitation spectrum of the insulating Bose glass is identified as the (intensive) energy scale at which  excitations become nearly delocalized. 
Within the forward scattering approximation in the bosonic hopping we find
the effective mobility edge $\epsilon_c$ to oscillate periodically with the magnetic
flux per plaquette, $\phi$. We find non-analytic cusps in $\epsilon_c(\phi)$ at integer or half-integer flux. 
The bosonic magneto-oscillations start with an increase of the mobility edge (and thus of resistance) with applied flux, in contrast to the equivalent fermionic problem. The amplitude of the
oscillations is much more substantial in bosons than in fermions. Bosons exhibit a single hump  per flux period, while fermion characteristics undergo two humps. Those are identical for non-interacting fermions, but Coulomb correlations are shown to lead to systematic deviations from this statistical period doubling.  Our theory reproduces  several key features observed in the activated magneto-transport in structured films.
 \end{abstract}
\maketitle
\enlargethispage{2\baselineskip}


\section{Introduction}
 
The interplay between disorder and Coulomb interactions is a crucial element affecting the phenomenology of the superconductor-insulator quantum phase transition. If only disorder and local BCS attraction is considered, and Coulomb repulsion is neglected, numerous theoretical studies \cite{MaLee,Kap,Fis90,Gho,Fei07,Fei10, Kra12,Bur12} have predicted the existence of preformed pairs in the vicinity of criticality, in the sense that the route from the insulating to the superconducting state proceeds directly through a delocalization of attractively bound pairs of electrons. This contrasts with the fermionic scenario first studied by Finkel'stein, in which the transition is driven by the suppression of electron pairing  due to disorder-enhanced Coulomb interactions~\cite{Fin87}. 
Under certain circumstances and in specific materials, however, it has been argued that the {\em local} Coulomb repulsion can be overcompensated by specific attraction mechanisms, resulting in systems with effective negative Hubbard $U$ interactions \cite{Gho01, FeiIM10,-UInO,-UPbTe}. 

On the experimental side, in the early nineties, transport measurements on InO$_x$ by Hebard, Palaanen, and Ruel \cite{Heb90, Paa92} were interpreted as signatures of Cooper pair insulators, suggesting that the above bosonic mechanism might be at work in that material \cite{Fis90}. Indeed, fermionic and bosonic insulators differ qualitatively since the exchange statistics affect their localization properties, in particular the interference of scattering paths that determine the decay of the wavefunction. In the presence of a magnetic field,  the wavefunctions of fermions and bosons respond in opposite ways~\cite{MM13,Syz12,Gang13,NSS,SS91,IS13}. For low energy bosonic excitations, the constructive interference among all paths is suppressed by a magnetic field, which leads to a strong positive magnetoresistance.~\cite{MM13} This contrasts with the subtle mechanism of the field-induced suppression of occasional negative interferences, which dominates the localization properties of localized fermions and results in a negative, but rather weak magnetoresistance.~\cite{NSS}

More recent experiments on amorphous thin films of Bi \cite{JV1,JV2, Goldman14}, PbBi \cite{JV14}, InO$_x$ \cite{Sam04,Kop12,Ste05}, TiN \cite{ Bat07}, or on a single ring of InO$_x$ \cite{Gurovich15} have strengthened the case of bosonic insulators, and exhibited a variety of intriguing transport characteristics. In particular, transport in the insulating state was observed to have an activated characteristics, with an Arrhenius-type resistance of the form $R(T)\propto \exp(T_0/T)$, over a significant range of temperatures, $T_0$ being the activation energy~\cite{Sam04, Shahar16}. 
Patterned films with an artificially created superlattice~\cite{JV1,JV2} also exhibited activated behavior, with an activation energy oscillating with the applied magnetic field. The observed oscillation period corresponds to one superconducting flux quantum $h/2e$ threading the unit cell of the superlattice, suggesting that the relevant charge carriers are pairs of electrons, which preserve phase  coherence beyond the scale of the imposed pattern.

The observation of purely activated transport in these systems is rather surprising in a highly disordered insulator, where generically a stretched exponential dependence of the resistance on temperature is expected, due to variable range hopping transport~\cite{MM09}. The latter, relies however, on a sufficiently efficient bath that allows inelastic transitions of  carriers to transport charge through the system. 
If instead the coupling between phonons and the relevant carriers (pairs or electrons) is weak, and if the low energy sector of electronic excitations is by itself discrete in nature, transport may be dominated by other channels than phonon-assisted variable range hopping. One possibility is the transport via activation to a mobility edge of the relevant charge carriers \cite{MM09, Iof10}, which indeed yields an Arrhenius resistance down to relatively low temperatures until eventually variable range hopping will take over, in spite of the inefficiency of the phonon bath. Such a phenomenology may be seen as a precursor of the much more stringent many-body localization, which not only requires a strong decoupling from phonons, but also the full localization of {\em any} intensive excitations, and in particular the absence of finite-energy mobility edges, which we discuss here. 

The above mentioned Arrhenius resistance is also expected in a wide temperature range if the mobility of charge excitations  merely exhibits a sharp crossover around an 'effective mobility edge' (in energy), instead of undergoing a genuinely sharp transition from localized to diffusive behavior at a precise energy.~\cite{MM09} This will be discussed in more detail below. 

In this work we explore the phenomenology of the crossover from weak to strong localization.
In particular we ask, how the effective mobility edge behaves in the presence of a magnetic field. At a qualitative level, it is clear that the effective mobility edge follow  trends analogous to those predicted for the localization length of low energy excitations: As the localization length increases, the effective mobility edge decreases, and vice versa~\cite{MM13}. Here we investigate this effect more quantitatively and show that a relatively simple model of  strongly localized pairs, subject to long range Coulomb interactions, is able to reproduce the salient features reported in the experiments on patterned films.   
  
Long range Coulomb interactions are known to play an important role in disordered insulators. In particular, they induce a depletion of the density of states around the chemical potential, creating a pseudo gap in the single particle density of states~\cite{Efr75}. This in turn modifies the localization properties of low energy excitations and promotes the appearance of an effective mobility edge, as was recently analyzed in the context of interacting electrons close to the Anderson-Mott metal insulator transition~\cite{Epperlein97, Ami14,Bur13}.
In contrast, in the presence of a flat or featureless bare density of states, with purely local repulsive interactions, there is no clear evidence of a mobility edge in the low energy spectrum of bosonic or fermionic insulators~\cite{MM13, Yu13}.
 Rather, the available techniques suggest that the localization length always decreases with increasing excitation energy. However, numerical results suggest that the addition of interactions, which are not strictly local,  induces a delocalizing tendency at higher energies, and thus mobility edges~\cite{Cue12}. The latter tendency becomes stronger with an increasing range of the interactions.
 Here we analyze the experimentally relevant case of unscreened, long range Coulomb interactions, and study the effect of magnetic fields on the effective mobility edge. Under the assumption that the effective mobility edge  takes the role of the activation energy $T_0$ that enters an Arrhenius law of transport, we obtain a semiquantitative description of transport in the absence of an efficient thermal bath.  

It is a main goal of this work to contrast the magnetoresistance in bosonic and fermionic systems. A particularly clean case can be made by comparing tightly bound pairs, acting as hard core bosons, with unpaired (spinless) fermions, which otherwise are subject to the same potential disorder, interactions and hopping strengths. Indeed, both carriers are hard core particles. The only difference consists in their exchange statistics, which at first sight might seem rather innocuous in insulators. However, they reflect strongly in the magnetoresistance, which probes the quantum interference in the exponential tails of localized excitations.

The remainder of this paper is organized as follows. In Sec.~\ref{model} we introduce and motivate the model under study. The magneto-oscillations of the localization length and the effective mobility edge for bosons are presented in detail in Sec.~\ref{boson}. Sec.~\ref{experiment} establishes the connection of our theory with experimental data. In Sec.~\ref{BvsF} we contrast the phenomenology of hard core bosons with that of fermions and explain the various effects of quantum statistics on the effective mobility edge. A summary of the central results is given in Sec.~\ref{summary}.


\section {Model}
\label{model}
\begin{figure}[!ht]
\includegraphics[width =0.35\textwidth ]{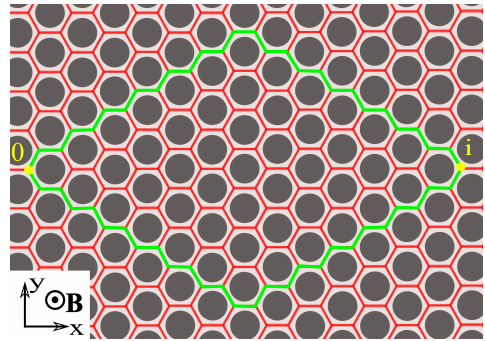}
\caption{Sketch of barely percolating films, with a triangular lattice of holes pinching it. These structures are modelled by a honeycomb lattice of islands hosting preformed pairs. The green lines connecting the two sites $0$ and $i$ enclose a diamond-shaped region containing all the shortest paths that connect those sites. }
\label{fig:sketch}
\end{figure}

The present study is motivated by the experiments of Refs.~ \onlinecite{JV1, JV2} on patterned films of Bismuth, with holes punched in a triangular  array. Those leave a  connected part of Bismuth  forming a 
 honeycomb lattice (with lattice constant $a\approx 50 {\rm nm}$), see Fig.~\ref{fig:sketch}. As those films are made sufficiently thin they undergo a superconductor-to-insulator transition, whereby the transport on the insulating side bears the hallmarks of a bosonic insulator. In particular, it exhibits a strong positive magnetoresistance. 
  
To model such films, we introduce a simplified model of interacting hard-core bosons \cite{MaLee} living on a two-dimensional honeycomb lattice of tunnel-coupled islands, governed by the Hamiltonian 
\bea
H &=&  \sum_i (\varepsilon_i - \mu) n_i  + \frac{1}{2}\sum_{j \neq i} \frac{q^2}{\kappa r_{ij}}\left(n_i - \nu\right)\left(n_j - \nu\right)\nn\\
 && - t\sum_{\langle ij \rangle}\left(\mathrm{e}^{\rm i \frac{q}{\hbar c}\int_{\mathbf{r}_j}^{\mathbf{r}_i}\mathbf{A}d \mathbf{r}}b_i^\dagger b _j + \mathrm{h.c} \right),
\label {Ham}
\eea
where $b_i^\dagger$, $b_i$ are the creation and annihilation operators of a hard-core boson of charge $q=2e$ on site $i$, and $n_i=b_i^\dagger b_i$ is the local number operator.  The hard core bosons represent strongly bound, preformed electron pairs. The chemical potential $\mu$ is adjusted such as to assure half-filling ($\nu=1/2$) of the lattice.
 The particles are subject to disordered onsite potentials $\varepsilon_i$ being uniformly distributed in $\varepsilon_i\in [-W,W]$.
They interact via long-range Coulomb interactions that decay as $1/r$, since the ambient space is 3d. $\kappa$ denotes the dielectric constant of the film, which is typically fairly large in such nearly metallic structures \cite{Fei10,Shk08}. The Coulomb contribution from a neutralizing background charge of homogeneous density $\nu$ has been subtracted.
The magnetic field enters via an Aharonov-Bohm phase factor multiplying the nearest-neighbor hopping amplitude $t$.  The phase acquired on each link is the line integral of the vector potential $\mathbf{A}$, for which we choose the gauge $\mathbf{A}=Bx\mathbf{e_y}$. We measure the magnetic field $B$ in terms of the fraction of flux quanta per plaquette, $f=B/B_0$, where  $B_0=hc/qS$, and $S= 3\sqrt{3} a^2/2$ is the area of the unit cell. The depairing Zeeman effect of the magnetic field  on the electron pairs is neglected here. Its effect will be studied in forthcoming work~\cite{Thuong2}.

The above model captures a rather generic situation in bosonic or spin-polarized fermionic insulators. Even though a given island $i$  will in general host  a rather large number of charges, in the insulating phase we may  restrict ourselves to describing the two most relevant charge states, which  differ by the absence or presence of a charge carrier (an electron pair in the case of the bosonic insulator). States differing by stronger charge fluctuations are not expected to modify the physical behavior of the insulator significantly, and thus we believe the above model to capture the gist of the experimental systems.

In the numerical studies carried out below, we study two-dimensional lattices and employ periodic boundary conditions. The Coulomb interaction between two sites is taken to be proportional to the inverse of the minimum distance on the torus. The Coulomb repulsion between nearest neighbor charges, $E_C=q^2/\kappa a$, is used as the unit of energy, while the lattice constant $a$ serves as the unit of length. 
 
\subsection{Classical electron pair glass}

\begin{figure}[!ht]
\includegraphics[width=0.43\textwidth]{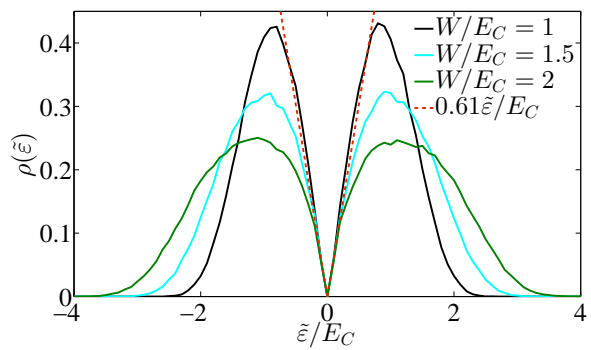}
\caption{Single-particle density of states for various disorder strengths $W$ in a two-dimensional Coulomb glass. A  linear Coulomb gap forms, which ensures stability with respect to single particle transitions. Excitations are more strongly localized at low energies. An effective mobility edge may thus appear at higher energies within the Coulomb gap.}
\label{f1}
\end{figure}

It is impossible to solve the full Hamiltonian (\ref{Ham}) exactly. Instead we approach the problem in an approximate way, which  captures the main physical effects. 
We consider the hopping as a perturbation and neglect it in a first step. That is, we first deal with a classical Hamiltonian describing a Coulomb glass of particles with charge $q=2e$. Such a system is well-known to possess many metastable low-energy configurations which are stable  with respect to the rearrangement of few particles. The Coulomb interactions with other particles strongly modify the distribution of the low-lying single-site excitation energies $\tilde{\varepsilon}_i$,
\bea
\tilde{\varepsilon}_i=\frac{dH}{dn_i}=  \varepsilon_i - \mu+ \sum_{j\neq i} \frac{q^2}{\kappa r_{ij}}(n_j - \nu).
\eea
In $d=2$ the Coulomb interactions create a linear Coulomb gap in the density of single particle excitations at low energy, 
$\rho(\tilde{\varepsilon}) = C \tilde{\varepsilon}/E_C^2$,
as predicted by Efros and Shklovskii \cite{Efr75}. Fig.~\ref{f1} shows the corresponding single-particle density of states, $\rho(\tilde{\varepsilon})$ for various disorder strengths, as obtained numerically. 
The coefficient $C$ is nearly independent of disorder (for $W\gtrsim 1$) and takes roughly the value $C\approx 0.61$, not far from theoretical predictions~\cite{Bar79}.

\subsection{Localization on the background of a Coulomb gap}
In the strongly insulating regime $t\ll W$, the hopping term can be treated as a perturbation. Here we study the localization properties of a single particle excitation. It can be read off from the spatial dependence of the amplitude of poles of the Green's function. Following the formalism introduced in Ref.~\onlinecite{MM13}, the Green's function (in the $T\to 0$ limit) can be obtained in a perturbative expansion in the hopping using a locator expansion, whereby we treat the onsite potentials $\tilde{\varepsilon}_i$ as frozen-in static disorder which depends on the metastable state under consideration. In a given metastable state of the Coulomb glass (defined by a locally stable classical charge distribution), to leading order in the hopping, the Green's function at large distance is obtained as
\bea
\frac{G_{0,i}(\omega,B)}{G_{0,0}(\omega,B)} &&= t^{r_{0i}} \sum_{
\substack{
   \Gamma: 0 \to i \\
   |\Gamma|=r_{0i}
  }
}
e^{i\Phi_\Gamma (B)} \prod_{k\in \Gamma\backslash \{0\}}\frac{{\rm sgn}(\tilde{\varepsilon}_k)}{\tilde{\varepsilon}_k-\omega}
\nn\\
&&\equiv \left(\frac{t}{W}\right)^{r_{0i}}  S_{0i}(\omega,B).
\label{GF} 
\eea
Here the sum $S_{0i}(\omega,B)$ runs over all paths $\Gamma$ of shortest length $|\Gamma|=r_{0i} \equiv {\rm dist}(0,i)$, defined as the minimal number of nearest neighbor hops necessary to connect the two sites. $\Phi_\Gamma(B)$ is the flux enclosed by the loop formed by path  $\Gamma$ and a fixed reference path connecting $0$ and $i$. The latter merely fixes the gauge of the Green's function.

In Eq.~(\ref{GF}), the only trace of quantum statistics is the residue ${\rm sgn}(\tilde{\varepsilon}_k)$ of the locator, which applies to hard core bosons. For non-interacting fermions, instead, this factor is absent.
This forward scattering approximation, and especially its fermionic version,  has been analyzed extensively in the literature~\cite{NSS, SS91, MM13, Gang13, Medina92, Kardar07}.

The localization length of excitations at energy $\varepsilon_0$ is defined as the inverse of the typical  spatial decay rate of  Green's function residues of poles at $\varepsilon=\varepsilon_0$,
\be
\label{xi_def}
\xi^{-1} (\varepsilon_0, B) = -\lim_{r_{0i}\to \infty} \frac{1}{r_{0i}}\overline{ \ln  \left| \frac{G_{0,i}(\omega, B)}{G_{0,0}(\omega,B)}\right|}_{\omega\rightarrow \varepsilon_0}.
\ee
The overbar denotes the disorder average. On a regular lattice, this definition depends on the direction in which the point $i$  tends to infinite distance from $0$, even though the relative variations will be very similar for different directions. Below we analyze the direction along a lattice base vector, as indicated in Fig.~\ref{fig:sketch}.

From Eq.~(\ref{GF}) it follows that at low excitation energies, $\omega \to 0$, in the absence of a  magnetic field ($\Phi=0$) all  paths come with positive amplitudes and thus interfere constructively. A magnetic field destroys the perfect constructive interference by adding a phase factor to each path. In contrast, for fermions, the path amplitudes always have essentially random signs, whatever the magnetic field. However, for $B=0$ the likelihood of occasional, strongly destructive interferences between two bunches of paths is bigger than in finite flux. This effect was first discovered by Nguyen, Spivak and Shklovskii.~\cite{NSS} It leads to a weak negative magnetoresistance for fermions, which contrasts with the strong positive response of bosons~\cite{Gang13}.

It is convenient to split the inverse localization length into a simple hopping part and a geometric  part capturing interference, 
\bea
\xi^{-1} (\varepsilon_0, B) = \ln\left(\frac{W}{t}\right) + \xi_g^{-1} (\varepsilon_0, B),
\eea
where
\bea
\xi_g^{-1} (\varepsilon_0, B) = -\lim_{r_{0i}\to \infty}
 \frac{1}{r_{0i}}\overline{ \ln  \left|  S_{0i}(\omega,B) \right|}_{\omega\rightarrow \varepsilon_0}.
\eea

\subsubsection*{Definition of (effective) mobility edge}
Due to the increase of the single particle density of states with energy $\varepsilon$, based on formula (\ref{GF}) one expects an increase of the localization length with increasing excitation energy $|\varepsilon-\mu|$. If the tunneling amplitude $t$ is finite, the localization length of zero temperature excitations, as defined by (\ref{xi_def}), may diverge at sufficiently high energies. This is indeed expected to happen in dimensions $d>2$ close enough to the transition to a conductor. This was analyzed in quite some detail for fermionic insulators in Refs.~ \onlinecite{Ami14} and ~\onlinecite{Bur13}.   
In such higher dimensional systems the energy
\bea
\epsilon_c={\rm inf}\{E| \xi(E)=\infty\}.
\label{EME}
 \eea
sharply defines a mobility edge in the limit $T\to 0$.
 
However, in dimensions $d=2$ (the case of interest to us here) at $T=0$, one  does not generally expect genuine delocalization at finite excitation energies. Rather, in close analogy with the well-known case of single particle excitations in the absence of anti-localizing spin-orbit interactions, one expects the proliferation of returns to the origin of any finite energy excitation to induce localization, albeit with a localization length that may become  exponentially large upon varying a control parameter. In non-interacting fermionic problems the control parameter is given by $k\ell $, which is to be considered as a function of the energy $E$. 

Nonetheless, even in $d=2$ it is meaningful to identify a crossover energy $\epsilon_c$ at which strong localization (at lower energies) turns into exponentially weak localization (at higher energies). For most practical purposes, such a crossover scale $\epsilon_c$ acts like an {\em effective} mobility edge, above which the effects of localization become very weak. They will thus not show up down to very low temperatures. If the localization length is a strongly increasing function of excitation energy the effective mobility edge is expected to exhibit only a slow logarithmic increase with decreasing temperature.
To illustrate this idea, let us briefly discuss the case of two-dimensional disordered insulators, where one expects that any finite energy excitation remains localized at strictly zero temperature. In other words, eigenstates with excitation energy $O(1)$ above the ground state are expected to differ only locally from the latter. One may in principle construct operators that create such "elementary" $T=0$ excitations from the ground state. However, in general two such operators do not (anti-)commute with each other. 
As a consequence,  eigenstates at finite energy density will not 
simply consist in a finite density of such localized excitations above the ground state, but hybridize various configurations with excitations in different locations. In particular the sufficiently weakly localized excitations at high energy will not commute (and thus collide) with many other elementary excitations. If the corresponding collision rate is bigger than the inverse of the level spacing in the localization volume of the high energy excitation, the localization of the latter should be irrelevant at that temperature, and one expects those excitations to be diffusive. This phenomenology leads to a weakly temperature dependent effective mobility edge, as was discussed in Ref. ~\onlinecite{MM09}. 
At sufficiently low temperature, the collision rate with other elementary excitations will eventually become so infrequent that the finite system size becomes a more efficient cut-off for localization. In that case the effective mobility edge will become (weakly) size dependent.~\cite{fn}

A practical definition for an effective mobility edge can be obtained by identifying the energy $\epsilon_c$ where the perturbative locator expansion (\ref{xi_def}) ceases to decay with distance (while higher order loop corrections would most likely reinstate a weak exponential decay), i.e., 
\bea
\epsilon_c={\rm min}\{E| \xi^{\rm FSA}(E)=\infty\}.
 \eea
Here, the superscript ${\rm FSA}$ indicates the restriction to the leading order forward scattering approximation.
 For non-interacting fermions in $d\geq 3$ this criterion correctly selects an energy for which $ k \ell(\epsilon_c) = O(1)$, a qualitative criterion which is also satisfied by the rigorously defined, sharp mobility edge (\ref{EME}). We stress that we are not so much interested in the absolute value of $\epsilon_c$ at a given set of parameters, but rather in its variations with magnetic field. We expect the qualitative features of such variations to be much less sensitive to the approximations involved in the restriction to forward scattering, than $\epsilon_c$ itself.
 
As mentioned before, in the absence of an efficient phonon or electron bath, the above defined $\epsilon_c$ will act like a mobility edge and may dominate transport 
in an intermediate temperature regime where activation to $\epsilon_c$ is less costly than weakly assisted variable range hopping passing through  lower lying energy states. Under such circumstances one may expect $\epsilon_c$ to appear as the activation energy in an Arrhenius-type resistance.~\cite{MM13, Fei10} 


\section {Bosonic localization: Oscillations of localization length and mobility edge}
\label{boson}
\subsection{Energy and field dependence of the localization length}

\begin{figure}[!ht]
      \includegraphics[width=0.42\textwidth]{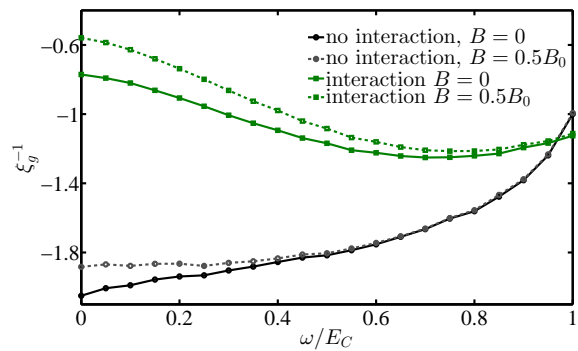}
\caption{Geometric part of the inverse localization length of hard core bosons as a function of excitation energy $\omega$. Without interactions and in the absence of a field, the localization length slightly decreases with increasing $\omega$.
The interaction-induced Coulomb gap enhances localization and reverses this trend, as localization becomes strongly enhanced at low energies. In either case the localization length shrinks with magnetic field (i.e., $\xi^{-1}_g$ increases). The effect is strongest at low energies, where the zero field interference is maximally constructive.}
\label{f2}
      \end{figure}

Fig.~\ref{f2} shows the numerically evaluated interference part of the inverse localization length as a function of excitation energy.
At $\omega=0$ all paths contribute  positively to a maximally constructive interference sum, while at finite energy occasional negative locators occur.
In the absence of interactions, i.e. without Coulomb gap in the density of states (data plotted in black), this leads to a slight increase of $\xi^{-1}_g$ with increasing $\omega$.~\cite{MM13, Yu13}
A magnetic field frustrates the predominantly positive interference and leads to a shrinkage of the localization length (positive magnetoresistance). This effect is  strongest for small $\omega$ where the field-free interference is maximal. 

Adding Coulomb interactions has quite a dramatic effect on the localization. The presence of the Coulomb gap suppresses the low energy density of states and thus strongly enhances the localization tendency there. The localization length qualitatively traces the variation of the density of states. Hence, the enhancement of localization is the stronger the lower the energy.  This overcompensates the effect of rarer and rarer negative locators as $\omega \to 0$. Within the forward approximation, the Coulomb gap indeed turns $\xi(\omega)$ into an increasing function of $\omega$, even at $B=0$, unlike in the limit of purely local hard core repulsions. 

If the hopping is sufficiently strong, high energy excitations are essentially delocalized and there is an effective mobility edge, as defined in (\ref{EME}). A magnetic field frustrates the predominantly constructive interference. This makes the localization length at a given energy shrink and thus pushes up the effective mobility edge.

\begin{figure}[!h]
\includegraphics[width=0.45\textwidth]{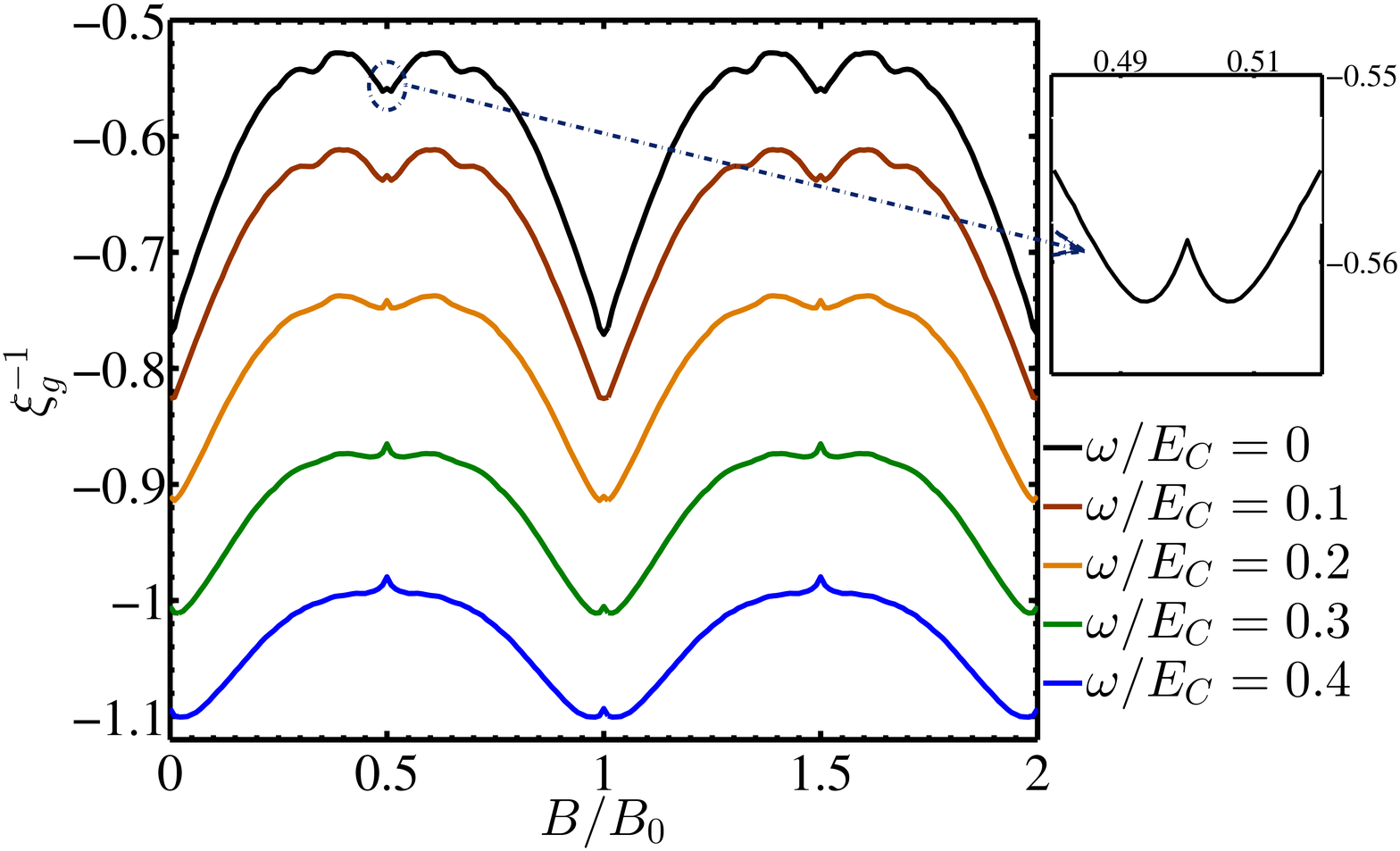}
\caption{Geometric part of the inverse localization length of bosonic excitations as a function of magnetic field, at various excitation energies. $\xi_g^{-1}$ tends toward a local minimum as the flux approaches integer values, or fractions with small denominators. There, for low energies, large subsets of paths interfere maximally positively. At finite energies, a tiny, non-analytic upward cusp of $\xi_g^{-1}$ sits on top of this main feature. It reflects the destruction of negative interference at large scales, akin to the dominant mechanism of magnetoresistance in fermions. Similar cusps of the same origin appear at half integer fluxes, cf. the inset (for $\omega=0$). } 
\label{f3}
\end{figure} 
\vspace{1em}
Fig.~\ref{f3} presents the full flux dependence of the inverse localization length. Its geometric part $\xi_g^{-1}$  oscillates with the period of one flux quantum per plaquette, $B_0$. At $\omega=0$ and for small fields, $B\ll B_0$, the localization length shrinks monotonically with increasing flux.
However, at finite excitation energies the localization length is slightly non-monotonic very close to $B=0$, even though this is hard to see in Fig.~\ref{f3} except at larger $\omega\gtrsim 0.2$.
Indeed, at non-zero energies locators occasionally have negative signs. At large scales the interference sum thus behaves like a fermionic problem, having a negative magnetoresistance at the smallest fields. This argument assumes the absence of the so-called sign transition, as discussed, e.g., in Refs. ~\onlinecite{Medina92} or ~\onlinecite{Huse11}. 
A small $B$-field then first reduces the destructive interference of paths with opposite signs, like in fermions, resulting in a very weak increase of $\xi$. A larger flux, however, has the main effect of suppressing the predominantly positive interference between shorter path segments. This then turns the magnetoresistance positive. This non-monotonicity in $\xi(B)$, which occurs for a small enough abundance of negative locators (i.e., not too large $\omega$), was already observed and explained in Ref.~\onlinecite{SS91} (cf. especially Fig.~3.2).

At half integer flux, $B=B_0/2$, further features appear in $\xi(B)$. At that flux all path amplitudes are real, but they fluctuate in sign. At exactly half-integer flux, the localization length is a local minimum of $\xi(B)$. This is reflected in a tiny upward cusp in $\xi_g^{-1}$, as illustrated by the inset of Fig.~\ref{f3}. It originates again from the elimination of occasional destructive interferences once the flux per plaquette deviates slightly from half integer. However, at larger deviations the dominant effect of $B$ is the destruction of maximal interference between  paths that differ by two unit cells; at least for sufficiently low energies $\omega$ where negative locators are rare. This results in an increase of $\xi_g^{-1}$. Similar local minima can be seen at the lowest $\omega$ for fluxes that are multiples of $B_0/3$. 

The cusps at integer and half-integer flux are all non-analytic. This can be understood from a mapping to directed polymers. The mapping is truly faithful at $\omega=0$, where all path weights are positive.~\cite{Gang13} However, also negative weight problems exhibit the same type of  scaling for the spatial  roughness of paths (with wandering exponent $\zeta=2/3$ in $d=2$), and amplitude fluctuations governed by a Tracy-Widom distribution~\cite{IS13,Prior09}. From those, one predicts a change of the localization length which scales as $\delta \xi_g^{-1} \sim |\delta B|^{\psi}$ with the deviation $\delta B$ from integer or half-integer flux, where the exponent has the value $\psi = 2\zeta/(1+\zeta) = 4/5$. ~\cite{Gang13}

\subsection{Magneto-oscillation of the effective mobility edge}
\begin{figure}[!ht]
\includegraphics[width=3.1in]{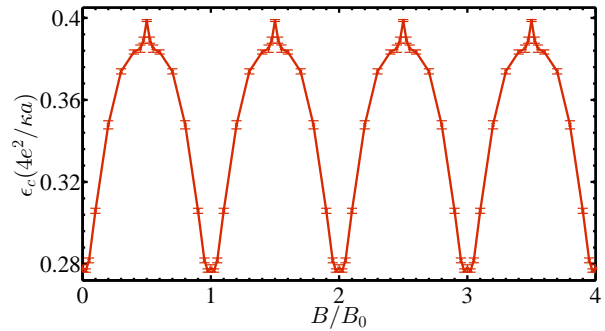}
\caption{Flux dependence of the effective mobility edge of bosonic excitations. 
The upward cusps $\sim |\delta B|^{4/5}$ at half integer fluxes, and similar (but tiny) cusps at integer fluxes originate from the destruction of occasional negative interference among certain close pairs of paths with real amplitudes but opposite signs. The overall dome shape of the oscillation reflects the destruction of the predominantly positive interference by the Aharonov-Bohm phases introduced by incommensurate fluxes.}
 \label{f4}
\end{figure}

For energies well inside the Coulomb gap, the localization length $\xi(\omega)$ is a monotonically growing function of $\omega$. For  sufficiently large hopping amplitude $t$, $\xi$ diverges at the finite effective mobility edge  $\epsilon_c$, which is a periodic function of the flux.  In Fig.~\ref{f4} we plot $\epsilon_c(B)$ for a fixed value of the hopping amplitude, $t = 0.368E_C$, and disorder strength $W= E_C$. 
With these parameters, we find the amplitude of oscillations of $\epsilon_c$  to be about   $\Delta \epsilon_c \approx 0.1 E_C$. 
The qualitative features of the field dependence $\epsilon_c(B)$ are the same as those of $\xi^{-1}_g(B, \omega)$ (cf. Fig.~\ref{f3}) for an energy $\omega \approx 0.3 E_C$ corresponding to the flux-averaged average mobility edge.
Upon approaching criticality, as the average mobility edge decreases, we expect the function $\epsilon_c(B)$ to become non-monotonic in the range $B\in [0,B_0/2]$, exhibiting maxima slightly before and after $B_0/2$, in analogy to the field dependence of $\xi_g^{-1}$ at low energies, cf. Fig.~\ref{f3}. However, we do not show  corresponding results of the forward scattering analysis, since so close to criticality our approximation is for sure not reliable quantitatively; even though the discussed qualitative features presumably survive.

Qualitatively, $\epsilon_c(B)$ shows the same features as those of $\xi(B,\omega>0)$. After a tiny, non-analytical decrease at $B\ll B_0$, the effective mobility edge increases as a consequence of the suppressed constructive interference in low energy bosonic excitations. 
At half flux per plaquette, $\epsilon_c(B)$ exhibits an upward cusp ~ $|\delta B|^{4/5}$, like $\xi_g^{-1}(B)$. Its origin lies in the destruction of occasional, nearly complete negative interferences.   

\subsection{Increased relative oscillations upon approach to criticality}
Note that as long as the effective mobility edge lies well within the Coulomb gap $\epsilon_c \lesssim E_{\rm gap} = E_C^2/(2CW)$ the disorder strength $W$ plays a minor role, since the smallest locators have an abundance dictated by the pseudo-gapped part of the density of states, which is nearly disorder independent. 

\begin{figure}[!ht]
\includegraphics[width=2.8in]{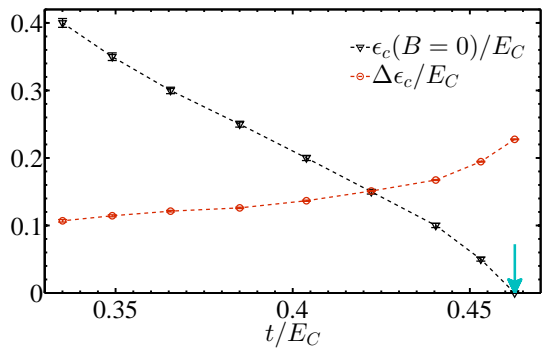}
\caption{The zero-field effective mobility edge  $\epsilon_c(B=0)$ and the magneto-oscillation amplitude $\Delta \epsilon_c$, plotted as a function of the hopping amplitude $t$. The mobility edge $\epsilon_c$ can be tuned by the hopping $t$. It serves as a measure for the distance to criticality. As the mobility edge $\epsilon_c$ decreases and the transition is approached (approximately where $\epsilon_c (B=0)\approx 0$, as marked by the arrow),  the oscillation amplitude increases.}
 \label{fig:osc_vs_t}
\end{figure}

In contrast, the hopping amplitude $t$ affects the location of the effective mobility edge directly, as illustrated in Fig.~\ref{fig:osc_vs_t}. That figure  shows that, upon tuning the hopping between islands, the oscillation amplitude increases as the effective mobility edge decreases, i.e.,  as the transition to the superconductor is approached. The location of the transition can roughly be estimated from the criterion $\epsilon_c(B=0)\approx 0$, but in its vicinity the forward scattering approximation should not be trusted quantitatively. For some range beyond the zero-field transition, the magnetic field is expected to be able to drive an SI transition.

It is interesting to compare these qualitative predictions with experimental data. To do so we interpret $\epsilon_c$ as the activation energy entering the Arrhenius-type resistance, and $\Delta \epsilon_c$ its field-induced variation.
The experiments of Refs. ~\onlinecite{JV1} (Fig. 3) and ~\onlinecite{JV14} (Fig. 3(b)) show the same trends as we find from our theory: the further  the system is from criticality, the smaller is the variation of the activation energy.

\section{Relating theory to experiments}
\label{experiment}
The experimental structured films~\cite{JV1,JV2,Kop12} bear signatures of bosonic insulators, the small field magnetoresistance being positive, while the flux periodicity is that expected for charges $q=2e$. We note that unpaired, non-interacting electrons of charge $q=1e$ would exhibit the same flux periodicity as we recall in the next section; however, 
as we will discuss there, in the presence of interactions the period of single electrons is doubled and thus faithfully reflects the carrier's charge.
 
To relate our theoretical study to experimental systems, we need to discuss the relevant scale of Coulomb interactions, $E_C$.
For an insulator of  bosonic  carriers of charge $q=2e$, with a lattice spacing between islands $a\approx 50{\rm nm}$ and dielectric constant $\kappa$ one obtains the Coulomb scale $E_C = q^2/\kappa a \approx 1334/\kappa\, {\rm K}$.  The essential difficulty resides in determining the effective dielectric constant $\kappa$ which governs the Coulomb interaction at and above the lattice scale $a$. This is nearly impossible to predict from first principles as the islands  possess a large polarizability and have to be considered as nearly touching each other. Therefore they renormalize  the dielectric constant of the medium surrounding the patterned film, such that values of $\kappa \sim 10^2 - 10^3$ are not unrealistic.   

However, another consideration allows us to argue for an upper bound on $E_C$, simply on empirical grounds.
The system essentially realizes an array of Josephson junctions. The proximity to the superconductor suggests that the charging energy ($\sim E_C$) is of the order of the Josephson energy, whose role is played by the hopping $t$ here. Deeply in the superconducting phase, the Josephson coupling determines the scale of the transition temperature $T_c$. These considerations imply that not too far from criticality $E_C $ is of the order of typical $T_c$ in well superconducting samples. Empirically, the latter never exceeds a few Kelvin, suggesting that  $E_C\sim 2 {\rm K}$, and effectively $\kappa \sim 500$.  

Our results in Fig.~\ref{fig:osc_vs_t} show that typical magneto-oscillation amplitudes are of the order of one magnitude smaller than $E_C$. This is compatible with experimental oscillation amplitudes of activation energies of the order of $0.2K$, as extracted from resistance data that were fitted to an Arrhenius law.~\cite{JV2}

Our theory  predicts a non-analytic cusp of the effective mobility edge at half integer fluxes, and another cusp of much smaller size at integer flux.
Interestingly, such cuspy features have  been observed in measurements of the resistance as a function of $B$, cf. Ref. ~\onlinecite{JV1},  Fig. 2A. 

As we discussed in the previous section,
we further expect that upon approaching criticality, when $\epsilon_c \lesssim 0.1 E_C$, the resistance develops a double-hump within an oscillation period, akin to the low energy behavior of $\xi_g^{-1}(\omega)$. Unfortunately, in the experimental systems of Refs. ~\onlinecite{JV1} and ~\onlinecite{JV2} this corresponds to a rather small energy scale. Therefore very low temperatures will be  requireed to reliably observe an activated behavior over a sufficient range of resistances and extract activation energies from it that would exhibit this double-hump feature.

\section {Role of quantum statistics - Bosonic vs fermionic mobility edges}
\label{BvsF}
\begin{figure}[!ht]
\includegraphics[width=0.43\textwidth]{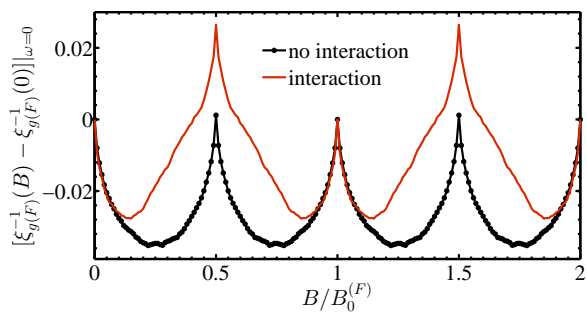}
\caption{Variations of the inverse localization length $\xi_{g,(F)}^{-1}$ of fermionic excitations at $\omega=0$  as a function of magnetic field - with and without interactions. In the non-interacting case, the symmetry in the distribution of the {\em uncorrelated} disorder potential leads to a doubling of the oscillation period. In the presence of interactions, the effective disorder is correlated, which re-instates the flux periodicity expected for fermions, $B_0^{(F)}= hc/e$. 
The correlations due to Coulomb repulsion enhance the localization at half a flux per plaquette as compared to commensurate flux, as explained in the main text. 
}
\label{f5}
\end{figure}

\begin{figure}[!ht]
\includegraphics[width=0.42\textwidth]{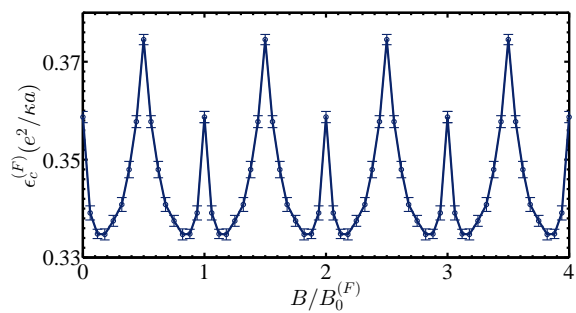}
\caption{Magnetic field dependence of the fermionic effective mobility edge $\epsilon_C^{(F)}$. The qualitative features are  similar to the variation of the inverse localization length in Fig.~\ref{f5}.} 
\label{f6}
\end{figure}

Apart from studying bosonic insulators per se, a central goal of this study is to investigate the role of quantum statistics in insulators. To this end we repeated the same type of analysis as above for a system of spinless fermions, subject to the same Coulomb interactions. The only difference with respect to the previously considered hard core bosons consists in the exchange statistics of the particles, while the Hilbert space and the terms in the Hamiltonian were left essentially identical. Data for the inverse localization lengths and effective mobility edges of fermions are shown in Figs.~\ref{f5} and \ref{f6}. The effective mobility edge of fermions oscillates with magnetic flux similarly as $\xi^{-1}$ at finite $\omega$.

The comparison between Figs.~\ref{f3} and \ref{f5} show three main effects of the opposite exchange statistics, some of which have been discussed previously in the literature~\cite{SS91,Gang13,IS13}: (i) the magnetoresistance of fermions in small fields is opposite to that of bosons at low energies; (ii) the amplitude of the field-induced variations are significantly smaller in fermions; (iii) the structure within an oscillation period is very different: bosons show one dome shaped oscillation, whereas fermions exhibit a pronounced double hump with a second local maximum in the localization length at half flux. As we discuss below the details of the latter reflect the nature of Coulomb correlations. Let us now explain these features in turn.
  
\subsection{Negative magnetoresistance of fermions}  
The increase of the fermion's localization length at small fields, as opposed to the stronger decrease in low energy bosons, is due to the fact that at $B=0$ fermionic paths already  come with random signs, so that there is no dominant positive interference to be destroyed by an extra $B$-field. Instead it is the $B$-induced lifting of accidental negative interference between two bunches of paths of nearly equal amplitude, which dominates the magnetoresistance by occasionally enhancing the tunneling further away. Such negative interferences are not that abundant, however. Therefore the resulting negative magnetoresistance is significantly less strong than the suppression of maximally positive interference of all bosonic paths. This explains the smaller amplitude of the field-induced variations in fermions.~\cite{Gang13}  

Fermionic path sums also obey the scaling of the Kardar-Parisi-Zhang universality class~\cite{Prior09}. Probabilistic arguments on the occurrence of large, strongly interfering pairs of path bundles~\cite{Gang13,IS13} thus lead again to the prediction that $\xi^{-1}$, as well as the effective mobility edge, vary in a non-analytical fashion close to integer and half integer fluxes as $\delta \xi^{-1} \sim  -|\delta B |^{4/5}$. 

\subsection{Approximate period doubling and traces of interaction correlations in fermionic magneto-oscillations}
An interesting, hitherto little explored feature is the structure of the magneto-oscillation within a flux period. For fermions there are two local maxima of $\xi^{-1}$ within one period. They occur at integer and half integer flux, where all path amplitudes are real (albeit random in sign). This maximally favors negative interference.
In fact, it has been known for a long time (cf., for example, Ref.~\onlinecite{SS91}, Fig.~3.2) that in non-interacting models, for an energy at the center of a symmetric impurity band, the magneto-oscillations of $\xi$ have a shorter period, reduced from $B_0$ to $B_0/2$, with identical peaks at integer and half integer flux, as we reconfirm in Fig.~\ref{f5}. For completeness, the proof of this fact is given in App.~\ref{app}. It relies on the symmetry of the distribution of onsite-potentials, $\rho(\omega+\delta)= \rho(\omega-\delta)$, and, most importantly, on the independence of potentials from site to site. 

The first assumption on the density of states is not that crucial. Indeed the deviations from perfect period doubling are not very significant as long as $\omega$ remains close to the band center of a featureless density of states. The  assumption of independence of onsite potentials is much more important. Crucially, it breaks down in the presence of interactions that induce correlations between local energies of spatially close sites. 
Indeed, around a soft site with a low local potential, non-local repulsive interactions suppress other sites with small potentials of opposite sign. That is, low energy sites in the vicinity of an occupied low energy site will predominantly be occupied themselves, rather than empty. Otherwise the considered configuration would be unstable with respect to the transfer from the occupied to the nearby empty sites. 

This bunching effect of  low energy  sites of the same kind has been described long ago in the literature of Coulomb glasses.~\cite{DLR84, BOR01} For the locator expansion in the insulating phase, it has the following interesting implication. Consider a small loop of interfering paths. Paths with significant weight contain a lot of small denominators, that is, they tend to pass through low energy sites. The correlation effect implies that two small denominators occuring in the two branches of a small loop are more likely to be of the same sign, and thus to interfere positively in the absence of flux. At the level of such a loop, adding half a flux through the plaquette is equivalent to flipping the sign of one of the energies. This induces a bias towards negatively interfering path pairs and thus enhances the localization tendency. The bias introduced by  correlations among nearby sites thus destroys the exact period doubling and induces maximal localization of fermions at half-integer flux, as confirmed by Fig.~\ref{f5}. 

Since this interaction effect is usually significantly stronger than the effect of a non-symmetric density of states, the deviation from period doubling in fermionic insulators can be used, qualitatively, as a measure and witness of Coulomb correlation effects.

\section{Summary and conclusion}
\label{summary}

\begin{figure}[!ht]
\includegraphics[width=0.4\textwidth]{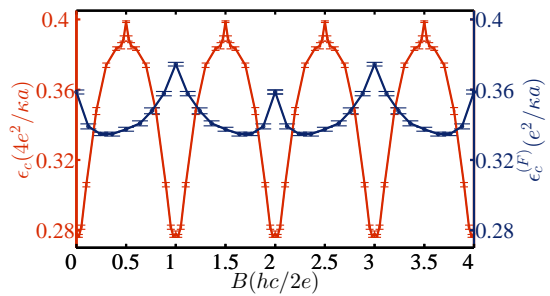}
\caption{Oscillations of the effective mobility edge of hardcore bosons of charge $2e$ versus that of  fermions of charge $e$. Each set of data is shown in  units of the relevant Coulomb interaction between nearest neighbors. Due to the approximate period doubling for fermions, the flux interval between maxima is the same as for bosons, but the structure within the oscillation period is very different: Fermions start with negative magnetoresistance at small fields, exhibit a smaller oscillation amplitude and alternating peak heights.}
\label{f7}
\end{figure}

In Fig.~\ref{f7} we provide a direct comparison of the oscillations of the effective mobility edge as a function of magnetic field for fermions of charge $e$ and those of hard core bosons (tightly bound electron pairs) of charge $2e$. 
Since these two systems share the same flux interval between peaks of enhanced localization, the latter cannot be used to determine the nature of the charge carriers. However, bosons and fermions are clearly distinguished by their opposite magnetoresistance close to integer fluxes: Bosons (at $\omega=0$) have a minimum of localization tendency at those points, whereas fermions exhibit a (weaker) maximum; a cousin of that fermionic maximum also appears at half integer flux. Note that the oscillation amplitude of the fermionic effective mobility edge is nearly one order of magnitude smaller than that of the bosons. 

As we explained in the last section, the correlations induced by repulsive interactions render the two fermionic maxima within a flux period inequivalent and enhance localization at half integer fluxes. We hope that future experiments on patterned films of non-superconducting metals will reveal these qualitative features reflecting both fermionic statistics and  correlations in the Coulomb glass.

Many aspects of our simple theoretical modelling are in reasonable semi-quantitative agreement with experimental data reported by J. Valles' group \cite{JV1, JV2, JV14}: The overall sign and shape of the magneto-oscillations, their cuspy nature at half flux as well as the evolution of their relative size as one tunes the distance to criticality. It would be interesting to test further predictions of our model, such as the appearance of a double hump in the oscillation period, as one approaches criticality more closely.

{\em Acknowledgment:} 
T. T. N. and M. M. acknowledge the hospitality of the University of Basel where part of this work was done.  T. T. N.  thanks the Paul Scherrer Institute for hospitality during an extended stay which allowed to complete this project.

\appendix

\section{Period doubling in the magnetoresistance of non-interacting fermions}
\label{app}

\begin{figure}[!ht]
\includegraphics[width=0.38\textwidth]{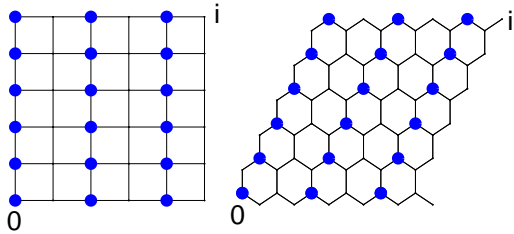}
\caption{Analyzing fermionic localization on two different lattices. For localization at energy $\omega=0$, adding half a flux quantum per unit cell is equivalent to having no flux and changing the sign of onsite disorder on the subset of  sites marked by circles, which yields a statistically equiprobable disorder configuration. For a symmetric disorder distribution, this property implies a period doubling of the magnetoresistance for non-interacting fermions, when evaluated in forward scattering approximation.}
\label{hex}
\end{figure}

This appendix recalls  the period-doubling in the magnetoresistance of non-interacting fermions on regular lattices, as evaluated within the forward scattering approximation. If the disordered onsite energies are uncorrelated and symmetrically distributed around $\omega=0$, one can prove that the localization length as a function of flux, $\xi(B)$, is a periodic function of $B$ with the reduced period $B_0/2$, $B_0$ corresponding to one flux quantum threading a unit cell of the lattice.  We show this for the cases of square and honeycomb lattices, see Fig.~\ref{hex}. In both lattices we marked a fraction of the sites with blue spots.

Consider the sum over shortest paths $\Gamma$ connecting site $0$ to site $i$, in the presence of a magnetic field $B$. 
Adding half a flux quantum per plaquette, one easily checks that the extra Aharonov-Bohm phase between two paths $\Gamma$ and $\Gamma'$ is given by $(-1)^{N_s}$, where $N_s$ is the number of marked sites that are not shared by both paths. One can verify that the same relative phase is obtained if the signs of all locators on the marked sites is reversed. 
This implies that up to a global sign the sum over paths at $\omega=0$ is equivalent to a sum in a field $B+B_0/2$, but with reversed sign of the onsite energy on marked sites. This change of sign leaves the measure of uncorrelated random energies invariant, provided the disorder distribution is symmetric around $\omega=0$, $\rho(\epsilon)= \rho(-\epsilon)$. From this, one concludes that $\xi(B) = \xi(B+B_0/2)$ is periodic with period $B_0/2$ for a symmetric disorder distribution and for $\omega=0$. 

For featureless densities of states and energies in the bulk of the spectrum the doubling of the periodicity is not exact, but nevertheless holds to a very good approximation. 

Note that time reversal symmetry further implies the symmetry $\xi(B) = \xi(-B)$. 

As we discuss in the main text, the above proof breaks down when the onsite energies are correlated, even if the density of states remains symmetric.



\begin{thebibliography}{99}
\bibitem{MaLee} M. Ma and P. A. Lee, Phys. Rev. B \textbf{32}, 5658 (1985). 

\bibitem{Kap} A. Kapitulnik and G. Kotliar, Phys. Rev. Lett. \tb{54}, 473 (1985); G. Kotliar and A. Kapitulnik, Phys. Rev. B \tb{33}, 3146 (1986).

\bibitem{Fis90} M. P. A. Fisher, Phys. Rev. Lett. \tb{65}, 923 (1990).

\bibitem{Gho} A. Ghosal, M. Randeria, and N. Trivedi, Phys. Rev. Lett. \tb{81}, 3940 (1998).

\bibitem{Fei07} M. V. Feigel'man, L. B. Ioffe, V. E. Kravtsov and E. A. Yuzbashyan, Phys. Rev. Lett. \textbf{98}, 027001 (2007). 

\bibitem{Fei10} M. V. Feigel'man, L. B. Ioffe, V. E. Kravtsov and E. Cuevas, Ann. Phy. \textbf{325}, 1390 (2010). 

\bibitem{Kra12} V. E. Kravtsov, J. Phys.: Conf. Ser. \tb{376}, 012003 (2012).

\bibitem{Bur12} I. S. Burmistrov, I. V. Gornyi, and A. D. Mirlin, Phys. Rev. Lett. \tb{108}, 017002 (2012).

\bibitem{Fin87} A. M. Finkel'stein, JETP Lett. \tb{45}, 46 (1987). 

\bibitem{Gho01} A. Ghosal, M. Randeria, and N. Trivedi, Phy. Rev. B \tb{65}, 014501 (2001).
\bibitem{FeiIM10} M. V. Feigel'man, L. B. Ioffe, and M. M\'ezard,  Phy. Rev. B \tb{82}, 184534 (2010).

\bibitem{-UInO} P. Reunchan, X. Zhou, S. Limpijumnong, A. Janotti, and C. G. Van de Walle, Curr. Appl. Phys. \tb{11}, 296 (2011).

\bibitem{-UPbTe} M. Dzero and J. Schmalian, Phys. Rev. Lett. \tb{94}, 157003 (2005).

\bibitem{Heb90} A. F. Hebard, M. A. Paalanen, Phys. Rev. Lett. \tb{65}, 927 (1990).
\bibitem{Paa92} M. A. Paalanen, A. F. Hebard, and R. R. Ruel, Phys. Rev. Lett. \tb{69}, 1604 (1992).

\bibitem{MM13} M. M\"uller, Europhys. Lett. \tb{102}, 67008 (2013).

\bibitem{Syz12} S. V. Syzranov, A. Moor, and K. B. Efetov, Phys. Rev. Lett. \tb{108}, 256601 (2012).

\bibitem{Gang13} A. Gangopadhyay, V. Galitski, M. M\"uller, Phys. Rev. Lett. 
\textbf{111}, 026801 (2013).

\bibitem{NSS}  V. L. Nguyen, B. Z. Spivak, B. I. Shklovskii, Pis'ma Zh. Eksp. Teor. Fiz. \tb{41}, 35 (1985) (JETP Lett. \tb{41}, 42 (1985)); Zh. Eksp. Teor. Fiz. \tb{89} 1770 (1985) (Sov. Phys. JETP \tb{62} 1021 (1985)).

\bibitem{SS91}  B. I. Shklovskii and B. Z. Spivak, in \textit{Hopping Transport in Solids}, edited by M. Pollak and B. Shklovskii (North-Holland, Amsterdam, 1991)

\bibitem{IS13}  L. B. Ioffe and B. Z. Spivak, JETP \tb{117}, 551 (2013).

\bibitem{JV1} \label{JV1} M. D. Stewart, Jr., A. Yin, J. M. Xu, and J. M. Valles, Jr., Science \textbf{318}, 1273 (2007).

\bibitem{JV2} H. Q. Nguyen, S. M. Hollen, M. D. Stewart, Jr., J. Shainline, A. Yin, J. M. Xu, and J. M. Valles, Jr., Phys. Rev. Lett. \textbf{103}, 157001 (2009).

\bibitem{Goldman14} Y.-H. Lin, J. Nelson, and A. M. Goldman, Phys. C: Superconduct. \tb{497}, 102 (2014).

\bibitem{JV14} S. M. Hollen, G. E. Fernandes, J. M. Xu, and J. M. Valles Phys. Rev. B \tb{90} 140506 (2014).

\bibitem{Sam04}  G. Sambandamurthy, L. W. Engel, A. Johansson, and D. Shahar, Phys. Rev. Lett. \tb{92}, 107005 (2004).


\bibitem{Kop12} G. Kopnov, O. Cohen, M. Ovadia, K. Hong Lee, C. C. Wong, and D. Shahar, Phys. Rev. Lett. \tb{109}, 167002 (2012).

\bibitem{Ste05} M. A. Steiner and A. Kapitulnik, Physica C \tb{422}, 16 (2005).


\bibitem{Bat07} T. I. Baturina , A. Y. Mironov, V. M. Vinokur, M. R. Baklanov, and C. Strunk, Phys. Rev. Lett. \tb{99}, 257003 (2007). 

\bibitem{Gurovich15} Doron Gurovich, Konstantin S. Tikhonov, Diana Mahalu, and Dan Shahar, Phys. Rev. B \tb{91}, 174505 (2015).

\bibitem{Shahar16} S. Mitra, G. C. Tewari, D. Mahalu, and D. Shahar, Phys. Rev. B \tb{93}, 155408 (2016).


\bibitem{MM09} M. M\"uller, Ann. Phys. (Berlin) \tb{18}, 849 (2009).
\bibitem{Iof10} L. B. Ioffe and M. M\'ezard, Phys. Rev. Lett. \tb{105}, 037001 (2010).

\bibitem{Fistul2008} M. V. Fistul, V. M. Vinokur, and T. I. Baturina, Phys. Rev. Lett. \tb{100}, 086805 (2008).
\bibitem{Ovadyahu2007} D. Kowal, Z. Ovadyahu, Physica C \tb{468}, 322 (2008).

\bibitem{fn} A simple example is given by single particle excitations in a non-interacting system with white-noise disorder in the continuum. The localization length $\xi(E)$ in the orthogonal universality class grows exponentially with energy, $\xi(E)\sim \exp[\gamma E]$. Transport through a finite  system of size $L$ then proceeds via  levels of energy $E$ that optimize the product $\exp[-E/T] \exp[-L/\xi(E)]$, which for non-interacting 2d electrons leads to a quasi-activated conduction with an activation energy that grows logarithmically with system size. Incidentally, activated transport with logarithmically growing activation energy was reported in insulating, bosonic systems in Ref. ~\onlinecite{Fistul2008}. We caution though that the scenario we mention is just one out of many possible explanations for such a phenomenology; it could possibly apply only under the stringent condition that the coupling to phonons and the ensuing variable range hopping transport are too weak to provide a more efficient transport channel in the considered temperature window. Simpler scenarii yielding similar length dependent insulating transport have been discussed in Ref. ~\onlinecite{Ovadyahu2007}.

\bibitem{Efr75} A. L. Efros and B. I. Shklovskii, J. Phys. C \tb{8}, 49 (1975); A. L. Efros and B. I. Shklovskii, \textit{Electronic Properties of Doped Semiconductors} (Springer, Berlin, 1984).


\bibitem{Epperlein97} F. Epperlein, M. Schreiber, and T. Vojta, Phys. Rev. B \tb{56}, 5890 (1997). 

\bibitem{Ami14} M. Amini, V. E. Kravtsov, and M. M\"uller, New J. Phys. \tb{16}, 015022 (2014).
\bibitem{Bur13} I. S. Burmistrov, I. V. Gornyi, and A. D. Mirlin, Phys. Rev. Lett. \tb{111}, 066601 (2013).

\bibitem{Yu13} X. Yu and M. M\"uller, Ann. Phys. (NY) \tb{337}, 55 (2013).


\bibitem{Cue12} E. Cuevas, M. Feigel'man, L. Ioffe, and M. M\'ezard, Nat. Commun. \tb{3}, 1128 (2012).

\bibitem{Shk08} B. I. Shklovskii, arXiv:0803.3331; A. I. Larkin, D. E. Khmelnitskii, Sov. Phys. JETP \textbf{56}, 647 (1982).

\bibitem{Thuong2} Thuong T. Nguyen and M. M\"uller, unpublished. 

 \bibitem{Bar79} S. D. Baranovskii, A. L. Efros, B. L. Gelmont, and B. I. Shklovskii, J. Phys. C: Solid State Phys. \tb{12}, 1023 (1979)

\bibitem{Medina92} E. Medina and M. Kardar, Phys. Rev. B \tb{46}, 9984 (1992). 

\bibitem{Kardar07} M. Kardar, \textit{Statistical Mechanics of Fields } (Cambridge University Press, Cambridge, England, 2007).

\bibitem{Huse11} H. Kim and D. A. Huse, Phys. Rev. B \tb{83}, 052405 (2011).

\bibitem{Prior09} J. Prior, A. M. Somoza, and M. Ortuno, Eur. Phys. J. B \tb{70}, 513 (2009).

\bibitem{DLR84} J. H. Davies, P. A. Lee, and T. M. Rice, Phys. Rev. B \tb{29}, 4260 (1984).

\bibitem{BOR01} S. A. Basylko, V. A. Onischouk, A. Rosengren, Phys. Rev. B \tb{65}, 024206 (2001).

\end{thebibliography}
\end{document}